\documentclass[12pt]{article}
\usepackage{epsf}

\begin{document}

\renewcommand{\theequation}{\arabic{section}.\arabic{equation}} 
\renewcommand{\thefootnote}{\fnsymbol{footnote}}                
\newfont{\elevenmib}{cmmib10 scaled\magstep1}
\newcommand{\preprint}{                                         
            \begin{flushleft}                                   
            \elevenmib Yukawa\, Institute\, Kyoto               
            \end{flushleft}\vspace{-1.3cm}                      
            \begin{flushright}\normalsize  \sf                  
            Preprint YITP-96-54 \\                              
            quant-ph/9610024 \\October 1996                     
            \end{flushright}}                                   
\newcommand{\Title}[1]{{\baselineskip=26pt \begin{center}       
            \Large \bf #1 \\ \ \\ \end{center}}}                
\newcommand{\Author}{\begin{center}\large \bf                   
            Hong-Chen Fu\footnote[1]{On leave of absence from   
            Institute of Theoretical Physics, Northeast         
            Normal University, Changchun 130024, P.R.China.     
            E-mail: hcfu@yukawa.kyoto-u.ac.jp }\ \              
            and Ryu Sasaki \end{center}}                        
\newcommand{\Address}{\begin{center} \it                        
            Yukawa Institute for Theoretical Physics, Kyoto     
            University,\\ Kyoto 606-01, Japan \end{center}}     
\newcommand{\Accepted}[1]{\begin{center}{\large \sf #1}\\       
            \vspace{1mm}{\small \sf Accepted for Publication}   
            \end{center}}                                       
\baselineskip=20pt
\preprint\bigskip
\Title{  Negative Binomial States of Quantized Radiation Fields}

\Author
\Address

\vspace{3.3cm}

\begin{abstract}
We introduce the {\it negative binomial states} with negative 
binomial distribution as their photon number distribution. They 
reduce to the ordinary coherent states and Susskind-Glogower phase
states in different limits. The ladder and displacement operator 
formalisms are found and they are essentially the Perelomov's 
$su(1,1)$ coherent states via its Holstein-Primakoff realisation. 
These states exhibit strong squeezing effect and they obey the 
super-Poissonian statistics. A method to generate these states is 
proposed.
\end{abstract}

\newpage

\section{Introduction}
\setcounter{equation}{0}

Nonclassical states of the radiation fields, such as the number
states, the coherent states and the phase states, play important
roles in quantum optics and are extensively studied \cite{noch}.
The binomial states (BS) introduced by Stoler {\it et al.} in
1985 \cite{stol}, interpolate between the {\it most nonclassical}
number states and the {\it most classical} coherent states, and
reduce to them in two different limits. Some of their properties
\cite{stol,leee,barr}, methods of generation \cite{stol,leee,datt},
as well as their interaction with atoms \cite{josh}, have been
investigated in the literature.
The notion of BS was also generalised to the intermediate
number-squeezed states \cite{bas1,fus4} and the number-phase states
\cite{bas2}, the hypergeometric states \cite{fus5},
as well as their $q$-deformation \cite{fann}.

The photon number distribution of BS is the binomial distribution
in the probability theory \cite{Feller,fus7}. In this letter,
we shall introduce and study {\it negative binomial states} (NBS)
whose photon distribution is the {\it negative binomial 
distribution} (NBD) \cite{Feller,fus7}. Different from
the BS, the NBS are the {\it intermediate phase-coherent states}
in the sense that they reduce to the Susskind-Glogower (SG) phase
states \cite{sg} and coherent states in two different limits
(Sec.2). We also derive their ladder and displacement operator
formalisms and find that they are essentially the Perelomov's
$su(1,1)$ coherent states via its Holstein-Primakoff (HP)
realizations (Sec.3). The NBS exhibit strong squeezing effects,
but are not of sub-Poissonian statistics (Sec.4). A method to
generate these NBS is proposed in Sec.5 and a summary with
special emphasis on the comparison with BS is given in Sec.6.

\section{Negative binomial states and their limits}
\setcounter{equation}{0}

We define the NBS as
\begin{equation}
   |\eta e^{i\theta};M\rangle^-  =\sum_{n=0}^{\infty}
     \sqrt{B^-_n(\eta ;M )}e^{in\theta}\| n\rangle,
   \label{nbs}
\end{equation}
in which  $\{\| n\rangle\, | n=0,1,\ldots\}$ are the number states of
the single-mode radiation field
\begin{equation}
   [b,b^\dagger]=1,\quad b\|0\rangle=0,
   \quad \| n\rangle={(b^\dagger)^n\over\sqrt{n!}}\|0\rangle,
   \label{bnum}
\end{equation}
$M$ is a fixed positive integer, $\eta^2$ is the probability
satisfying $0<\eta^2<1$, and
\begin{equation}
   B^-_n(\eta ;M )={M+n-1 \choose
   n}\eta^{2n}(1-\eta^2)^M,\quad n=0,1,\ldots.
   \label{nbd1}
\end{equation}
The $B^-_n(\eta ;M )$ is called the NBD \cite{Feller} since it can
also be written as
\begin{equation}
   B^-_n(\eta ;M )=(1-\eta^2)^M{-M \choose
   n}(-\eta^2)^n,\quad n=0,1,\ldots,
   \label{nbd2}
\end{equation}
which has the similar form as the binomial distribution except
for the two minuses and that $n$ runs to infinity. The states 
(\ref{nbs}) are referred to as 
the NBS since their photon distribution 
$|\langle n\| \eta e^{i\theta};M\rangle^-|^2 \equiv B^-_n(\eta ;M )$
is the NBD.

The parameter $\theta$ ($0\leq \theta <2\pi$) has clear 
physical meaning: it reflects
the time development of the NBS. This can be seen from 
$e^{-iHt}|\eta e^{i\theta};M\rangle^-
=|\eta e^{i(\theta-\omega t)};M\rangle^-$, where $H=\omega
(N+1/2)$ is the Hamiltonian of the single mode radiation field.

As a probability distribution $B^-_n(\eta ;M )$ satisfies
\cite{Feller}
\begin{equation}
   \sum_{n=0}^{\infty} B^-_n(\eta ;M )=1,   \label{condition}
\end{equation}
which means that the NBS are normalized. 

Let us consider two limiting  cases:

(1). In the limit $\eta\to 0$, $B^-_n(\eta ;M )\to\delta_{n0}$
and thus the NBS go to the vacuum state.

(2). When $\eta\to 0$, $M\to\infty$ with fixed
finite $\eta^2M =\alpha^2$, the NBD goes to the Poisson
distribution $B^-_n(\eta ;M )\to e^{-\alpha^2} \alpha^{2n}/n!$
\cite{Feller}. Accordingly, the NBS degenerate to the
ordinary coherent states.

We know that the BS degenerate to the number state in a
certain limit \cite{fus4}. However the NBS do not maintain this
feature. Instead, the NBS tend to the SG phase states in
a certain limit. To achieve this, let us consider the
case $M=1$. In this case, $|\eta e^{i\theta};M\rangle^-$ is
simplified as
\begin{equation}
     |\eta e^{i\theta};1\rangle^-=\sqrt{1-\eta^2}
     \sum_{n=0}^{\infty}\eta^n\,
     e^{in\theta}\| n\rangle. \label{geom}
\end{equation}
The photon distribution 
$|\langle n\| \eta e^{i\theta};1\rangle^-|^2$
in this case is $(1-\eta^2)\eta^{2n}$, the
geometric distribution. For this reason we call
$|\eta e^{i\theta};1\rangle^-$ the geometric states.
One can easily verify that the geometric state
$|\eta e^{i\theta};1\rangle^-$ is the eigenstate of the SG
phase operator $E^-$ \cite{sg} with the eigenvalue
$\eta e^{i\theta}$
\begin{equation}
   E^-=\sum_{n=0}^{\infty} \| n\rangle\langle n+1\|,
\end{equation}
which is related to the annihilation operator $b$
through polar decomposition
$b=E^-\sqrt{N}$. Now, multiplying a constant
$1/\sqrt{2\pi(1-\eta^2)}$ to $|\eta e^{i\theta};1\rangle^-$
and then taking the limit $\eta\to 1$, we obtain the
{\it phase states}
\begin{equation}
   |\theta\rangle=\lim_{\eta\to 1}\frac{1}{
   \sqrt{2\pi(1-\eta^2)}}|\eta e^{i\theta};1\rangle^-=
   \frac{1}{\sqrt{2\pi}}\sum_{n=0}^{\infty}
   e^{in\theta}\| n\rangle, \ \ \
   E^-|\theta\rangle=e^{i\theta}|\theta\rangle.
\end{equation}
The $|\theta\rangle$ states are non-normalisable,
nonorthogonal, but resolve the identity
\begin{equation}
   \int_{-\pi}^{\pi}\mbox{d}\theta
   |\theta\rangle\langle \theta|=1. \label{phstate}
\end{equation}
The phase-state representation based on (\ref{phstate})
is a useful calculational tool \cite{phre}.

In this sense, we find that the NBS are the {\it intermediate
phase-coherent states}. This is an important feature of the 
NBS.

\section{Displacement and ladder operator formalisms}
\setcounter{equation}{0}

Let us recapitulate the displacement operator formalism of NBS
\cite{fus7}
using the {\it identity method} developed in a previous paper 
\cite{fus2}. To this end, let us rewrite NBS (\ref{nbs}) 
($\eta_C\equiv\eta e^{i\theta}$)
\begin{equation}
	|\eta_C ;M\rangle^-=(1-|\eta_C|^2)^{M\over2}\sum_{n=0}^\infty
	{\sqrt{M(M+1)\cdots(M+n-1)}\over n!} 
	(\eta_C )^n(b^\dagger)^n\| 0\rangle.
	\label{3366}
\end{equation}
Then, by making use of the following identity
\begin{equation}
   (b^{\dagger}g(N))^n\|0\rangle=(b^{\dagger})^n
   g(0)g(1)\cdots g(n-1)\|0\rangle, \mbox{\ with \ }
   g(N)\equiv \sqrt{M+N},\quad N=b^\dagger b.
   \label{nbinide}
\end{equation}
we can write Eq.(\ref{3366}) in the exponential form
\begin{eqnarray}
   &&   |\eta_C ;M\rangle^-=(1-|\eta_C|^2)^{M\over 2}
   \exp\left[\eta_C\,{\cal K}_+\right]
   \|0\rangle, \nonumber \\
   &&   {\cal K}_+=b^{\dagger}
   \sqrt{M+N}\equiv \sqrt{M-1+N} b^{\dagger}.  \label{dis1}
\end{eqnarray}
It is interesting that ${\cal K}_+$ along with 
\begin{equation}
   {\cal K}_-=({\cal K}_+)^{\dagger}
   \equiv\sqrt{M+N}\,b\equiv b\sqrt{M+N-1}, \ \ \ \
   {\cal K}_0=\frac{M}{2}+N
\end{equation}
generates the $su(1,1)$ algebra via its HP
realization with the Bargmann index $M/2$. By making use 
of the disentangling theorem of $su(1,1)$ algebra \cite{trua}
we arrive at the displacement operator formalism of NBS
\begin{equation}
    |\eta e^{i\theta} ;M\rangle^-=\exp\left[\zeta_C\,{\cal K}_+
            -\zeta_C^*\,{\cal K}_-\right]
   \|0\rangle,\quad
   \zeta_C=e^{i\theta}\hbox{arctanh}\,\eta.
   \label{dis2}
\end{equation}
Eq.(\ref{dis2}) is nothing but the Perelomov's coherent 
state of $su(1,1)$ via its HP realisation.

The Perelomov's coherent states admit the ladder operator
form \cite{fus3}. To see this, we differentiate both
(\ref{dis1}) and (\ref{dis2}) with respect to $|\zeta|$
and equate the results. We have
\begin{equation}
   \left[ e^{-i\theta}{\cal K}_--\eta^2 e^{i\theta}
   {\cal K}_+\right]|\eta e^{i\theta} ;M\rangle^-=
   M\eta |\eta e^{i\theta} ;M\rangle^-.\label{ladder}
\end{equation}
This ladder operator form is obviously compatible with the
limit results. In fact, in the limits $\eta\to 0$ and
$\eta\to 0$, $M\to \infty$ with $M\eta^2=\alpha^2$, 
Eq.(\ref{ladder}) reduces to
\begin{equation}
   b|0 ;M\rangle^-=0, \ \ \
   b|0 ;\infty\rangle^-=\alpha e^{i\theta}|0 ;\infty\rangle^-,
\end{equation}
which mean $|0 ;M\rangle^-$ and $|0 ;\infty\rangle^-$ are the
vacuum and coherent states respectively.

Finally we point out that the NBS can also be regarded as the
{\em density dependent annihilation operator} coherent states, 
namely,
\begin{equation}
   E^-_M |\eta e^{i\theta} ;M\rangle^-=e^{i\theta}\eta
   |\eta e^{i\theta} ;M\rangle^-,\ \ \
   E^-_M\equiv b\frac{1}{\sqrt{N+M-1}}
\end{equation}
when $M\geq 2$. The operator $E^-_1$ is not well-defined in the 
vacuum state. If we require 
\begin{equation}
   E^-_1 \|0\rangle = 0,
\end{equation}
then it is obvious that $E^-_1=E^-$. This connection of SG phase 
operator with Perelomov's coherent states with Bargmann index
$1/2$ was well-known.

\section{Nonclassical effects}
\setcounter{equation}{0}

\subsection{Photon statistics}

Let us first examine if the NBS is of sub-Poissonian statistics.
Using Eq.\,(\ref{condition}), it is easy to
calculate the averages $\langle N\rangle$, $\langle
N^2\rangle$ and the fluctuation $\langle\Delta N^2
\rangle$ of the photon number
\begin{equation}
   \langle N\rangle=\frac{M\eta^2}{1-\eta^2},\ \ \
   \langle N^2 \rangle = \frac{M\eta^2+M^2\eta^4}{(1-\eta^2)^2},
   \ \ \
   \langle \Delta N^2 \rangle = \frac{M\eta^2}{(1-\eta^2)^2}.
\end{equation}
Then Mandel's $Q$-factor characterising sub-Poissonian (if $Q<0$)
distribution is obtained as
\begin{equation}
  Q=\frac{\langle \Delta N^2\rangle -\langle N\rangle}
  {\langle N\rangle}=\frac{\eta^2}{1-\eta^2}>0.
\end{equation}
So the field in NBS is {\it super-Poissonian}, not sub-Poissonian,
except for the (vacuum and coherent state) limit $\eta\to 0$
($Q\to 0$, Poissonian statistics).

Since the occurrence of
antibunching and sub-Poissonian are concomitant for single and
time independent field, as in the present case, the field in NBS
is bunching, not antibunching. In fact, from the following
second-order correlation function $g^2(0)$
\begin{equation}
   g^2(0)=\frac{\langle b^{\dagger}b^{\dagger}bb \rangle}{
   \langle b^{\dagger}b \rangle^2}
   =1+\frac{1}{M}>1,
\end{equation}
we can arrive at the same conclusion.

\subsection{Squeezing effect}

Let us evaluate the variances  $\langle \Delta x^2\rangle=
\langle x^2\rangle-\langle x\rangle^2$ and $
\langle \Delta p^2\rangle=\langle p^2\rangle-\langle p\rangle^2$ of
the quadrature operators $x$ (coordinate) and $p$
(momentum) defined by
\begin{equation}
   x=\frac{1}{\sqrt{2}}(b^{\dagger}+b), \ \ \ \
   p=\frac{i}{\sqrt{2}}(b^{\dagger}-b).
   \label{coomom}
\end{equation}
From the following relation 
\begin{equation}
   b^k|\eta e^{i\theta} ;M\rangle^-=
   \left({\eta e^{i\theta}\over\sqrt{1-\eta^2}}
   \right)^k
   \sqrt{M(M+1)\cdots (M+k-1)}\, |\eta e^{i\theta} ;M+k\rangle^-,
\end{equation}
we have
\begin{eqnarray}
   \langle \Delta p^2\rangle &=& \frac{1}{2}
        -\frac{\eta^2(1-\eta^2)^M}{(M-1)!}\sum_{n=0}^{\infty} 
        \eta^{2n}
        \frac{(M+n-1)!}{n!}\sqrt{M+n}\left(\cos(2\theta)\sqrt{M+n+1}-
        \sqrt{M+n}\right) \nonumber \\
        && - 2\sin^2\theta \,\eta^2  (1-\eta^2)^{2M}\left[
        \sum_{n=0}^{\infty} \left(\begin{array}{c}M+n-1\\n
        \end{array}
        \right)\eta^{2n}\sqrt{M+n}\right]^2,  \label{555}\\
   \langle \Delta x^2\rangle &=& \frac{1}{2}+
        \frac{\eta^2 M}{1-\eta^2}+
        \frac{\cos(2\theta)\eta^2(1-\eta^2)^M}{(M-1)!}
        \sum_{n=0}^{\infty}\eta^{2n} \frac{(M+n-1)!}{n!}
        \sqrt{(M+n+1)(M+n)}                   \nonumber \\ &&
        -2\cos^2\theta\, \eta^2  (1-\eta^2)^{2M} \left[
        \sum_{n=0}^{\infty}\left(\begin{array}{c} M+n-1\\n\end{array}
        \right)\eta^{2n} \sqrt{M+n} \right]^2.
\end{eqnarray}
In the derivation of Eq.(\ref{555}), we have used the identity
\begin{eqnarray}
    \frac{M\eta^2}{1-\eta^2}&=&\frac{\eta^2(1-\eta^2)^M}{(M-1)!}\,
    \frac{M!}{(1-\eta^2)^{M+1}}=
    \frac{\eta^2(1-\eta^2)^M}{(M-1)!}\sum_{n=0}^{\infty}
    \eta^{2n}\frac{(M+n)!}{n!}  \nonumber \\ &=&
    \frac{\eta^2(1-\eta^2)^M}{(M-1)!}\sum_{n=0}^{\infty}
    \eta^{2n}\frac{(M+n-1)!}{n!}\sqrt{M+n}\sqrt{M+n}.
\end{eqnarray}

Let us analyse $\langle \Delta p^2\rangle$ in more detail.
First consider the case $\theta=0$ (the initial time). In
this case Eq.(\ref{555}) is simplified as
\begin{equation}
    \langle \Delta p^2\rangle=\frac{1}{2}-
    \frac{\eta^2(1-\eta^2)^M}{(M-1)!}\, \sum_{n=0}^{\infty}
    \eta^{2n} \frac{(M+n-1)!}{n!}\sqrt{M+n}\left(
    \sqrt{M+n+1}-\sqrt{M+n}\right).  \label{delp}
\end{equation}
Since every term in the infinite series in Eq.(\ref{delp})
is positive, the infinite sum is positive. So we always 
have $\langle \Delta p^2\rangle<1/2$. This means that the 
quadrature $p$ is squeezed. 
Fig.1 shows how $P\equiv\langle\Delta p^2\rangle$
depends on $\eta$ and $M$. We
have chosen $0\leq\eta^2\leq 0.99$. Those plots show that: 

(1). Dependence on $\eta^2$. It is found that the 
variance  $\langle\Delta p^2\rangle$ is an increasing function 
of $\eta^2$, namely, the larger $\eta^2$, the stronger the
squeezing effect. Note that $\langle\Delta p^2\rangle>0$ since
$\langle\Delta p^2\rangle=\langle p^2\rangle$ and $p^2$ is a
positive definite hermitian operator.

(2). Dependence on $M$. We have chosen $M$=1, 5, 50. The squeezing
of $p$ for larger $M$ is stronger than
that for small $M$. When $\eta^2$ is small (close to 0) or large
(close to 1), the difference is very small. While when $\eta^2$
is around $1/2$, the difference is larger. However, squeezing
is not so sensitive to $M$. The case $M=5$ and $M=50$ are almost
same as showed in Fig.1.

Furthermore, due to the uncertainty relation
$\langle\Delta x^2\rangle\langle\Delta p^2\rangle \geq 1/4$,
the variance $\langle \Delta x^2\rangle>1/2$ when $\theta=0$. 
So the quadrature $x$ is not squeezed. 

When $\theta\neq 0$, it is easy to see that $\langle\Delta
p^2\rangle$ is a $\pi$-periodic function with respect to  $\theta$ 
and symmetric with respect to $\theta=\pi/2$. To
investigate the effect of $\theta$ to  $\langle\Delta p^2\rangle$,
we plot the  $\langle\Delta p^2\rangle$ as a function of $\theta$
for different $\eta^2$ values in Fig.2 (we choose $M=1$ for
simplicity). We find that (1) $\langle\Delta p^2\rangle$ becomes
larger when $\theta$ creasing. It first reaches $1/2$ and then
reaches  the maximal value when $\theta=\pi/2$. Then it 
symmetrically decreases until $\theta=\pi$. In some region
around $\theta=\pi/2$ (dependent on $\eta^2$), the
squeezing disappears. (2) Small $\eta^2$ violates the squeezing
slightly while large $\eta^2$ destroys the squeezing strongly.

\section{Generation of NBS}
\setcounter{equation}{0}

The displacement operator formalism suggests that the NBS can
be generated by the nondegenerate parameter amplifier described
by the Hamiltonian \cite{wall}
\begin{equation}
   H=H_0+ \chi i (a_1^{\dagger}a_2^{\dagger} e^{-2i\omega t}-
     a_1a_2 e^{2i\omega t}),\ \ \ \
   H_0=\omega_1 a_1^{\dagger}a_1 +
       \omega_2 a_2^{\dagger}a_2,
\end{equation}
where $a_1$ and $\omega_1$ ($a_2$ and $\omega_2$) are the 
annihilation operator and frequency for the signal
(idler) mode. Frequencies $\omega_1$ and $\omega_2$ sum to the 
pump frequency, $2\omega=\omega_1+\omega_2$ . The coupling 
constant $\chi$ is proportional to the second-order 
susceptibility of the medium and to the amplitude of the pump.
The unitary time evolution operator in the interaction picture is
\begin{equation}
   U(t)=e^{iH_0 t}e^{\chi t(a_1^{\dagger}a_2^{\dagger}
        -a_1 a_2)} e^{-iH_0 t}.
\end{equation}
Suppose that the system is initially prepared in the
state $|0,M\rangle\equiv\|0\rangle$. Since the photons are 
created or annihilated in pairs, we can restrict ourselves in 
the subspace
\begin{equation}
   \|n\rangle\equiv |n, n+M\rangle, \ \ \
   n=0,1,2,\cdots,
\end{equation}
which is isomorphic to the single-mode Fock space.
Then at any time $t$ the system is in 
\begin{equation}
   U(t)\|0\rangle=\sum_{n=0}^{\infty}\left[\left(
   \begin{array}{c}n+M-1\\n\end{array}\right)
   (1-\tanh^2\chi t)^M (\tanh\chi t)^{2n}\right]^{1\over 2}
   e^{i2\omega n}\|n\rangle.
\end{equation}
Identifying $\eta=\tanh\chi t$ and $\theta=2\omega t$,
we obtain the NBS.

\section{Conclusion}
In this letter we have introduced the negative binomial states
and studied their nonclassical properties. The following
table shows some properties of NBS with the special emphasis on the
comparison with the binomial states:
\bigskip

\begin{tabular}{|l|l|l|}\hline \hline
                  & NBS                   & BS     \\ \hline
Limit States      & Coherent and SG phase states &
                    Coherent and number states \\ \hline
Related algebra   & $su(1,1)$               & $su(2)$ \\ \hline
Essence           & Perelomov's coherent states & Atomic coherent
states\\ \hline
Sub-Poissonian     & No                     & Unconditional \\ \hline
Antibunching      & No                     & Unconditional \\ \hline
Squeezing effect  & Unconditional for $p$, & Limited ranges of 
                                             parameters \\ 
($\theta=0$)      & no squeezing for $x$   & for $x$, no squeezing for 
                                             $p$\\ 
                                             \hline\hline
\end{tabular}

\bigskip
We see that the nonclassical properties of the NBS and BS are
complementary.

As further work we shall generalize the notations of NBS in this 
letter to the {\it negative multinomial states} with negative 
multinomial distribution as their photon distribution. This 
generalisation concerns the multi-mode radiation field and should 
exhibit some more fruitful nonclassical properties like 
correlation between different modes. It is also a good 
challenges to study the interaction of radiation field 
in the NBS with the atoms.

\section*{Acknowledgements}

This work is supported partially by the grant-in-aid for
Scientific Research, Priority Area 231 ``Infinite Analysis'',
Japan Ministry of Education. H.\,C.\,F is grateful to Japan
Society for Promotion of Science (JSPS) for the fellowship.
He is also supported in part by the National Science
Foundation of China.


\newpage

\begin{figure}
\centerline{\epsfxsize=10cm 
\epsfbox{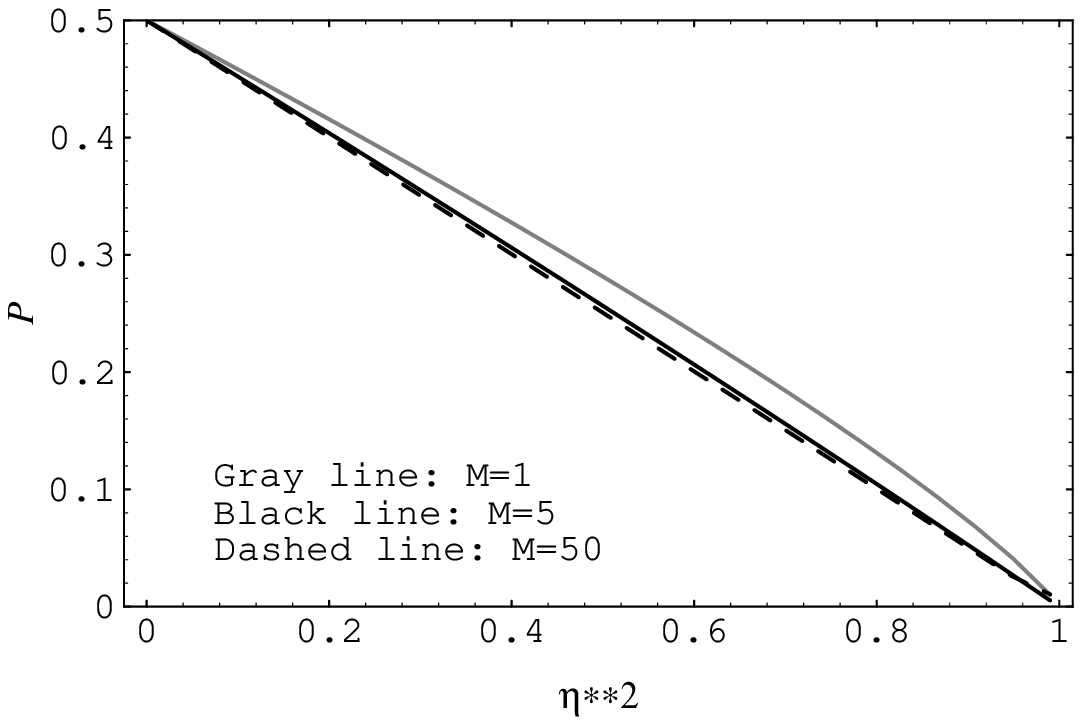}}
\caption{ Variance $P\equiv\langle\Delta p^2\rangle$ as a function
of $\eta^2\,(\equiv \eta **2)$ for $\theta=0$ (initial time) and $M$=1 (gray line), 5 
(black line) and 50 (dashed line).}
\end{figure}

\begin{figure}
\centerline{\epsfxsize=10cm 
\epsfbox{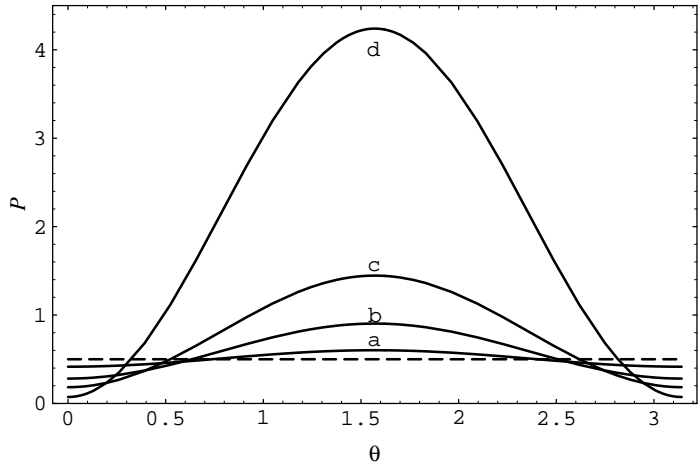}}
\caption{ Variance $P\equiv\langle\Delta p^2\rangle$ as a function
of $\theta$ for different $\eta^2$, 0.2 (a), 0.5 (b), 0.7 (c)
and 0.9 (d). In all the cases $M=1$. The dashed line corresponds to 
$\eta=0$ or $P=0.5$.}
\end{figure}

\end{document}